# Two-photon interference from independent cavity-coupled emitters on-a-chip

*Je-Hyung Kim,[†] Christopher J. K. Richardson,[‡] Richard P. Leavitt,[‡] and Edo Waks*[†,§]

[†]Department of Electrical and Computer Engineering and Institute for Research in Electronics and Applied Physics, University of Maryland, College Park, Maryland 20742, United States

[‡]Laboratory for Physical Sciences, University of Maryland, College Park, Maryland 20740, United States

[§]Joint Quantum Institute, University of Maryland and the National Institute of Standards and Technology, College Park, Maryland 20742 , United States

ABSTRACT. Interactions between solid-state quantum emitters and cavities are important for a broad range of applications in quantum communication, linear optical quantum computing, nonlinear photonics, and photonic quantum simulation. These applications often require combining many devices on a single chip with identical emission wavelengths in order to generate two-photon interference, the primary mechanism for achieving effective photon-photon interactions. Such integration remains extremely challenging due to inhomogeneous broadening and fabrication errors that randomize the resonant frequencies of both the emitters and cavities. In this letter we demonstrate two-photon interference from independent cavity-coupled emitters on the same chip, providing a potential solution to this long-standing problem. We overcome spectral mismatch between different cavities due to fabrication errors by depositing and locally evaporating a thin layer of condensed nitrogen. We integrate optical heaters to tune individual dots within each cavity to the same resonance with better than 3 μeV of precision. Combining these tuning methods, we demonstrate two-photon interference between two devices spaced by less than 15 μm on the same chip with a post-selected visibility of 33%. These results pave the way to integrate multiple quantum light sources on the same chip to develop quantum photonic devices.

Photonic quantum information processing requires efficient quantum light sources combined with optical networks that process photons.[1-4] Integrating these components onto a single chip would realize quantum photonic devices that enable quantum communication networks,[1,2] photonic quantum computers,[5,6] and photonic quantum simulators.[7-9] Many of these applications rely on optical qubits that exhibit two-photon quantum interference on a beamsplitter, the primary mechanism for achieving effective photon-photon interactions. A number of works have reported such two-photon interference from a variety of solid-state quantum emitters such as defect centers,[10,11] dophants,[12] and quantum dots.[13-22] But these sources exhibit isotropic emission that is often difficult to collect efficiently.

Optical cavities can significantly enhance the collection efficiency and spontaneous emission rate of quantum emitters.[22-25] Emitters coupled to cavities can also serve as highly nonlinear devices operating at low photon numbers,[26] as well as efficient interfaces between photons and solid-state quantum memory.[27,28] The majority of the work to-date focused on a single emitter in a cavity, which can exhibit nearly perfect single photon purity and indistinguishability using resonant pumping techniques.[14-17] Two-photon interference has been demonstrated from two cavity-coupled emitters on different chips contained in separate cryostats.[21] But integrating multiple cavity-coupled emitters on the same chip remains extremely challenging due to spectral randomness of the emitters and errors in nanofabrication. Spectral randomness destroys photon indistinguishability required for quantum information. Fabrication errors lead to spectral mismatch in the resonances of individual cavities, making it unlikely that multiple cavities will operate at the same frequency. Scalable quantum photonic devices require new methods to overcome both of these problems simultaneously.

In this letter we demonstrate a method to integrate multiple solid-state emitters resonantly coupled to cavities on the same chip, and show that these emitters exhibit two-photon interference. We couple multiple InAs/InP quantum dots that serve as efficient single photon sources[22,29,30] to independent photonic crystal cavities fabricated in close proximity on the same chip. In order to match the cavity resonances and compensate for fabrication errors, we utilize combination of nitrogen gas deposition and local thermal evaporation. To compensate for the inhomogeneous broadening of the quantum dot emitters, we fabricate thermally isolated photonic crystal devices integrated with optical heaters[31] that enable us to tune individual dots without affecting nearby devices. We combine these methods to match the resonant frequencies of independent quantum dots coupled to cavities separated by less than 15 μm on the same chip, and demonstrate that their emission exhibits two-photon interference. These results provide a path to integrate multiple quantum emitters and cavities on-a-chip for scalable quantum photonics applications.

Figure 1a shows a scanning electron microscopy image of a fabricated device composed of four *L3* photonic crystal cavities[32] with integrated optical heating pads[31] (See Methods for description of fabrication and design). We activate a particular heating pad by exciting it with a focused laser beam. The absorbed light heats the cavity attached to the pad. Each pad is thermally isolated from the substrate by 500 nm-wide thin tethers that suspend the membrane and provide poor thermal conductivity to the chip.

Figure 1b shows the experimental setup (see Methods for detailed description). We focus 780 nm laser beams on each cavity in order to excite the dots and create fluorescence emission. A third 780 nm laser beam thermally tunes one of the quantum dots by exciting the heating pad. The absorbed light does not affect the surrounding structures due to the poor thermal conductivity of the thin tethers[31]. We focus all laser beams using a single objective lens (NA=0.7) that also collects the emission from the devices using a confocal microscope geometry. A polarizing beamsplitter separates the collected emission of the horizontally oriented cavities from the vertically oriented cavities and injects them into opposite sides of a 50:50 beamsplitter to perform two-photon interference experiments. This configuration enables us to easily measure two-photon interference from two cavities with orthogonal orientation. We filter the mixed signal from the two cavities on a grating spectrometer and focus to a single mode fiber connected to InGaAs single photon counters.

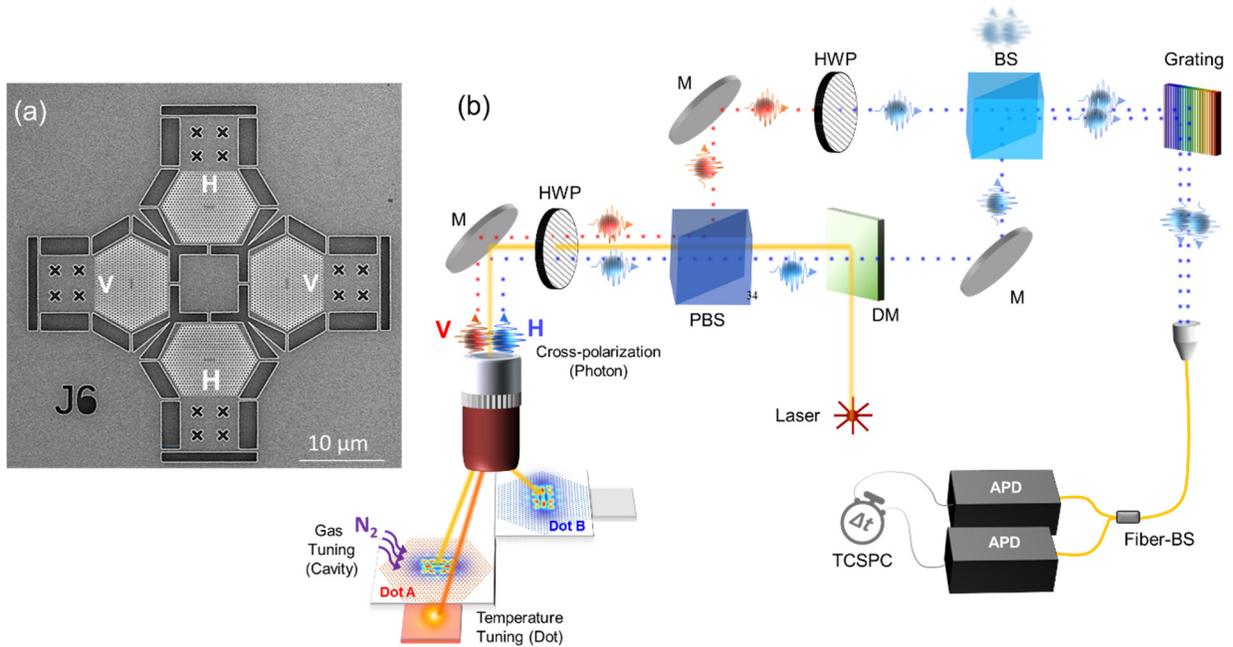

**Figure 1.** Sample design and experimental set-up. (a) Scanning electron microscopy image of a photonic crystal device. H and V denote horizontally and vertically oriented cavities. (b) Schematic of the experimental set-up for individual controls of cavity-coupled dots and two-photon interference. (M: mirror, DM: dichroic mirror, HWP: half-waveplate, BS: beamsplitter, PBS: polarizing beamsplitter, APD: avalanche photodiode, TCSPC: time-correlated single photon counting module).

Because we inject photons into a single mode fiber after the first collection lens, it is essential that emission from the cavity provide a Gaussian-shaped transverse mode. The photonic crystal cavity design we use has five prominent modes that we label M1-M5 consistent with previous notation.[22,33] The majority of works on photonic crystals and quantum dots use the lowest energy mode M1, which has the highest $Q$. But this mode has a transverse far-field profile that is difficult to collect and couple to a single mode fiber. Instead, we use mode M3 which has a transverse mode profile that is much closer to a Gaussian and can attain much higher collection efficiency into a single-mode fiber.[22,34]

Due to fabrication errors, the four cavities in the device shown in Figure 1a have different resonant frequencies for mode M3. To correct this spectral mismatch between the cavities we first inject a small quantity of nitrogen gas into the vacuum chamber. At 4 K the gas condenses onto the sample surface creating a thin layer that shifts the cavity resonance.[35] We tune the resonance of each cavity individually by locally heating the cavity using a focused laser beam and evaporating a fraction of the condensed gas layer. Figure 2a shows the cavity spectrum for one cycle of gas deposition and evaporation. The black line shows the original spectrum prior to injecting nitrogen gas into the vacuum. We observe two modes in the spectrum corresponding to modes M3 and M4 (all other modes reside outside the plotted spectral range). After injecting gas into the vacuum, both resonances red-shift by approximately 20 nm. We then focus a 780 nm continuous wave laser with an average power of 5 mW onto the cavity. Under this laser excitation the modes gradually blue shift due to evaporation of the condensed gas layer and return to their resonance frequencies after about 50 s. Below 5 mW of excitation power we do not observe any cavity mode shift indicating that this power is too low to cause evaporation. Thus, we can safely excite the quantum dot emission at low powers without inducing any shift to the cavity modes.

During the tuning process, we do not observe any shift in the adjacent cavities, indicating that we have attained sufficient thermal isolation.

Figure 2b and 2c show photoluminescence spectra from two adjacent cavities after we compensate for their spectral mismatch using local gas tuning. We plot the spectra using both a high excitation power of 1 mW (black line) and a low power of 1 μW (red line) (both of these powers are below the 5 mW threshold required to shift the cavity). The high power spectra from the two cavities show a resonance at identical wavelengths corresponding to the M3 cavity mode. At low power, we see bright sharp peaks within the cavity resonance corresponding to emissions from single quantum dots. One cavity has a sharp quantum dot resonance at an emission wavelength of 1250.40 nm, which we label as quantum dot A, while the second cavity has a quantum dot emitting at 1250.74 nm, which we label as quantum dot B. To verify that these resonances correspond to single quantum dots, we perform second-order autocorrelation measurements by using a 780 nm pulsed laser. Both dots exhibit anti-bunching at zero delay in Figure 2d,e, with a $g^{(2)}(0)$ of 0.25±0.01 and 0.29±0.01 for dots A and B respectively. The center peak in second-order correlation at zero time delay does not fully vanish due to background which is largely dominated by detector dark counts. When subtracting this dark count rate, which we independently measure, the corrected $g^{(2)}(0)$ is 0.076±0.017 for dot A and 0.12±0.012 for dot B (see Methods).

We observe an order of magnitude enhancement in the emission intensity of the cavity-coupled quantum dots compared to bulk dots, indicating that the cavity enhances the collection efficiency of the photon sources. From time-resolved lifetime measurements we obtain a lifetime of 1.12 ns and 1.06 ns for dots A and B respectively. For comparison, we perform lifetime measurements on quantum dots that are spectrally detuned from cavity modes by more than 5 nm. These dots show an average lifetime of 3.2±0.7 ns. From these measurements we can estimate the cavity coupling efficiency, $\beta = 1 - \frac{\tau_{on}}{\tau_{PhC}}$, where $\tau_{on}$ is the lifetime of dots A and B, and $\tau_{PhC} = 3.2$ ns is the lifetime of the spectrally detuned dots. From this equation we estimate $\beta \sim 0.65$, indicating that the cavity collects a large fraction of the emitted photons and re-directs them into a nearly Gaussian transverse mode. We estimate the collection efficiency into the first objective lens by exciting the device using a pulsed laser and comparing the detector count rates to the laser repetition rate. After accounting for the losses in the system (see Methods), we estimate a collection efficiency of 8% and 10% for dots A and B respectively.

We also characterize the polarization properties the emission dots A and B, which exhibits a high degree of linear polarization with polarization ratios of 0.96±0.01 and 0.93±0.01 (Figure 3). The polarization directions lie along the directions of the cavity polarizations, and we choose dots A and B in two adjacent cavities with orthogonal orientation so these dots produce cross-polarized emission. This polarized emission provides further evidence of high coupling efficiency and is also essential to separate the emission from two resonant dots using a polarizing beamsplitter, as shown in Figure 1b.

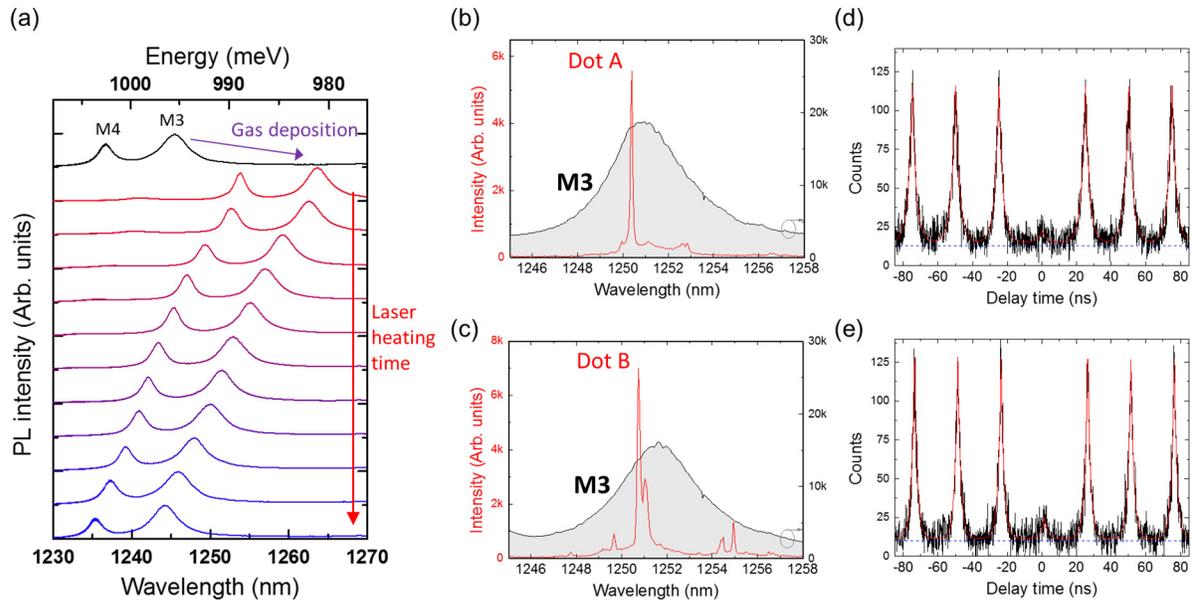

**Figure 2.** Cavity tuning and single photon emission from cavity-coupled dots (a) Cavity emission spectra after gas deposition and laser heating. The black line is the original spectrum prior to gas deposition, and red to blue-colored lines indicate the spectrum in time steps of 5s after laser heating. (b,c) Photoluminescence spectra from two individual cavities. The black line is the spectrum at an excitation power of 1 mW (right axis) and the red line is at an excitation power of 1 µW (left axis). (d,e) Pulsed second-order autocorrelation histograms for dots A and B. Red line is a fitted curve and blue dashed line is a background level due to detector dark counts.

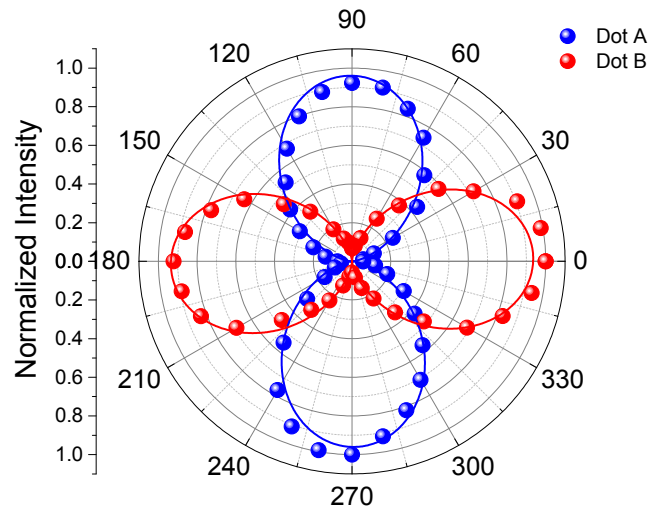

**Figure 3.** Polar plots of the single photon emission from dots A (blue dots) and B (red dots) as a function of polarization angle.

The resonance of dot A and B are detuned by 0.34 nm, which is much greater than the quantum dot linewidth. To compensate for this detuning we excite the heating pad connected to the device containing dot A with a 780 nm laser to tune it onto resonance with dot B. In addition to this heating laser, we also excite both dots with focused lasers to generate emission. We separate the orthogonal polarized emissions from the two cavities using a polarizing beamsplitter and re-combine them on a 50:50 beamsplitter, and send the combined signal to a spectrometer. Figure 4a shows the combined emission from both cavities as a function of the heating laser power. As we increase the power of the heating laser, the emission of dot A shows a red shift, while dot B remains unaffected due to thermal isolation between the cavities. Figure 4b shows the spectrum of each quantum dot at a heating laser power of 1.25 mW where the resonance frequencies are almost perfectly matched. By fitting the emission of the two dots to a Lorentzian, we determine that their resonances are within 3 µeV. After thermal tuning both dots still show a spectrometer-limited linewidth of 50 µeV. Thus, any linewidth broadening due to heating is below our instrumentation resolution. The heating laser creates additional background emission of 9% and 6% for dots A and B of the signal measured at single photon detectors, which partially degrades $g^{(2)}(0)$. We could eliminate this background by depositing metal on the heating pads[31] in order to efficiently absorb the incident light without emitting or by employing micro heaters.[36]

In order to determine the temperature that we are heating the sample, we perform a second experiment where we increase the temperature of the entire sample using an electrical heater connected to a sample stage (Figure 4c). As the temperature increases from 4 K to 25 K, both dots shift by 0.73 nm. By comparing the temperature of the sample stage that achieves the same shift as the 1.25 mW heating laser used to tune the dots onto resonance, we determine that the optical heater heats dot A to 16 K.

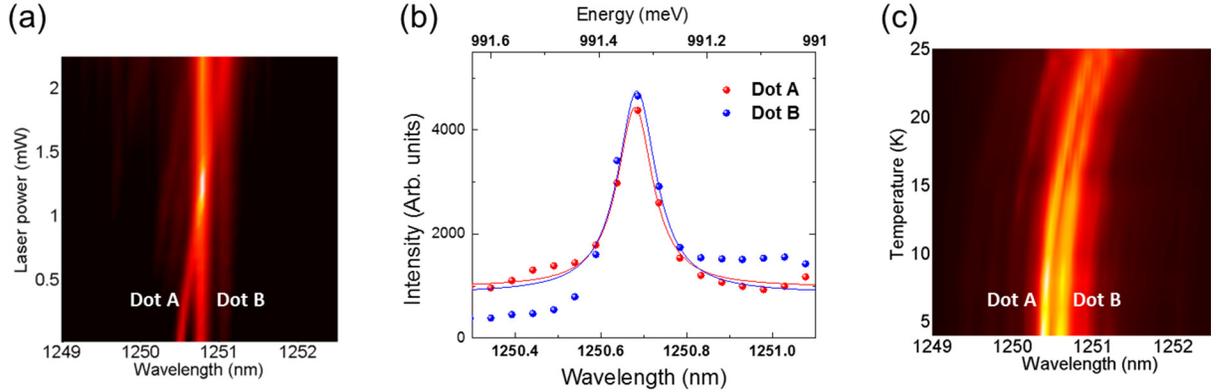

**Figure 4.** Thermal tuning of individual dots. (a) Emission spectra of dots A and B as a function of a laser power that locally heats dot A. (b) Photoluminescence spectra of dots A and B at a heating laser power of 1.25 mW. Solid curves are Lorentzian fits. (c) Emission spectra of dots A and B as a function of temperature of a sample stage.

Having matched both the cavity and quantum dot resonant frequency, we study two-photon interference of the emission from dots A and B. We use a modified version of a Hong-Ou-Mandel interferometer that places two detectors at the same side of the beamsplitter (Figure 1b).[37-39] For general Hong-Ou-Mandel experiments with two detectors at the opposite sides of the beamsplitter, destructive interference creates anti-bunching in the coincidence curves.[10,20] However, in this modified scheme, we collect the two-photon state from one arm of the path-entangled state $|\psi\rangle = (|2,0\rangle - |0,2\rangle)/\sqrt{2}$, and split it using a second 50:50 beamsplitter. The two-photon state results in bunching at zero delay time as opposed to anti-bunching.[39] We balance the emission intensities from each dot using a neutral density filter, and use a half-waveplate before one input arm of the 50:50 beamsplitter to control the relative polarization of the two photons in order to switch them from distinguishable to indistinguishable. A second fiber-based beamsplitter splits the emission from one arm of the 50:50 beamsplitter onto two InGaAs avalanche photodiodes to perform time-resolved coincidence detection. Delay time in the measured coincident counts denotes the time delay between photon detection on the two detectors. This approach has the practical advantage that we can use a single grating spectrometer to filter the photon emission from both dots, making the experimental setup significantly simpler. We perform the two-photon coincidence measurements at 50% of the saturation power of quantum dot emission.

Figure 5 shows the normalized coincidence histograms for indistinguishable (parallel polarization) and distinguishable (orthogonal polarization) single photons without background subtraction. Solid curve shows a theoretical fit that takes into account the finite time resolution of the detectors and the background emission due to the heating laser (See Methods). The orthogonally polarized photons show the expected anti-bunching behavior with a second order correlation at zero time delay of $g_\perp^{(2)}(0) = 0.72 \pm 0.01$, which is higher than the ideal value of 0.5 due to the limited time resolution, detector dark counts and background emission from the heating laser. Parallel-polarized photons show a sharp peak in the second-order correlation near zero delay time due to two-photon interference, where $g_\parallel^{(2)}(0) = 0.96 \pm 0.02$. The post-selected visibility of the two-photon interference effect is given by $V = (g_\parallel^{(2)}(0) - g_\perp^{(2)}(0))/g_\perp^{(2)}(0) = 0.33 \pm 0.01$.

The sharp peak in the indistinguishable two-photon correlation measurement indicates the presence of dephasing. From the width of the peak we can extract the coherence time of the quantum dot which is 115±5 ps. This coherence time is consistent with previous measurements performed at 4K,[22] and suggests that that thermal tuning does not significantly increase dephasing within our tuning range.[14] Rather than the pure dephasing at this temperature, we attribute the dominant dephasing mechanism in our system to timing jitter in the dot emission as a result of above-band excitation.[40] Despite this dephasing, the post-selected visibility at zero time delay should still attain unity in the ideal case, but in practice our measured visibility is limited by the finite time resolution of the detectors which is on the same order as the dephasing time, as well as background emission caused by the heating laser.

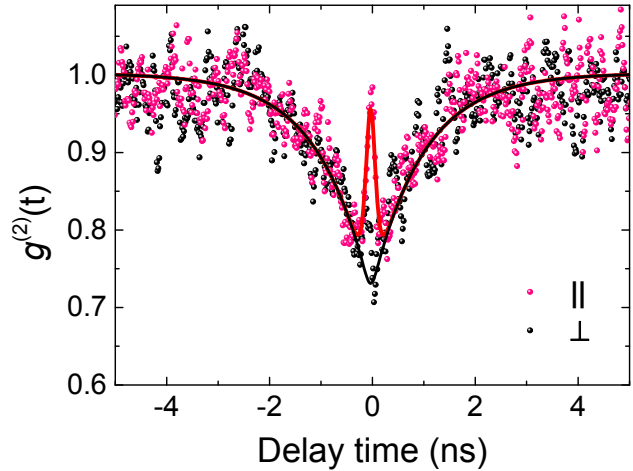

**Figure 5.** Normalized two-photon coincidence data for parallel (red dots) and orthogonal (black dots) polarized photons. Solid lines represent fitted curves based on a theoretical model.

In summary, we demonstrated two-photon interference from independent cavity-coupled dots on the same chip. We combined thermal tuning of quantum dots with nitrogen gas deposition and local evaporation to match the resonances of both cavities and dots. Using this approach we attained a two-photon interference visibility of 0.33. This interference demonstrates that we attain sufficient tuning range and precision to match individual dots separated in distance by less than 15 μm, a crucial requirement for scalable quantum photonics applications. These results represent an important step towards scalable quantum integrated photonic devices composed of multiple sources for photonic quantum information processing

One of the main limitations to our indistinguishability contrast is that we used non-resonant above-band excitation which can cause decoherence and large time jitter in the emitted photons[40]. We could significantly reduce this effect by using resonant pumping[15-17] or quasi-resonant pumping[41,42] The temperature tuning method we employ also provides only a limited tuning range because the linewidth of the quantum dots broadens with increasing temperature, which will ultimately degrade the two-photon interference contrast.[14] Future devices could incorporate DC Stark shift[20] or local strain tuning[43] to enable wider tuning ranges. InAs/InP quantum dots can also operate at longer wavelengths in the 1.55 μm range compatible with long-distance fiber communication. Finally, in our current experiment we generate photons on the same chip but measure two-photon interference off-chip. More scalable approaches would extract emission directly into a photonic crystal waveguide for on-chip processing of photons. A number of promising techniques can attain efficient extraction of light to a waveguide,[44-46] opening up the possibility to create complex circuits operating at the single photon level.

**Methods.**

*Sample information.* The initial sample contained InAs quantum dots with a density of approximately 10 μm$^{-2}$ in a 280 nm-thick InP membrane on a 2 μm-thick AlInAs sacrificial layer. The cavity design was based on an *L*3 defect cavity with a lattice parameter (denoted as *a*) of 350 nm, a hole radius of 0.27 *a*, and a slab thickness of 280 nm, and we shift the three holes adjacent to the cavity outward by 0.26 *a*, 0.16 *a*, and 0.16 *a*, respectively. To fabricate the device we deposited a 100 nm-thick silicon nitride thin film as an etching mask by using plasma-enhanced chemical vapor deposition. We then patterned the SiN layer using electron beam lithography and fluorine-based reactive ion etching, and transferred the pattern to the quantum dot sample using chlorine-based reactive ion etching. Finally, we removed the sacrificial layer by selective wet etching to form an air-suspended photonic crystal membrane.

*Experimental set-up.* We mounted the sample on a low-vibration closed cycle cryostat operating at 4 K. We used three 780 nm lasers to excited dots A and B and the heating pad for dot A, and controlled the focused beam positions of these lasers using several mirrors and beam splitters (not shown in Figure 1b). For photon counting experiments we used InGaAs avalanche single photon detectors with an efficiency of 20%, a dark count rate of 200 Hz, and a timing resolution of 200 ps. For pulsed $g^{(2)}(t)$ and lifetime measurements, we used a 780 nm pulsed laser diode with a pulse width of 50 ps and a repetition rate of 40 MHz. A time-correlated single-photon counting module recorded the correlation histograms with a time bin of 16 ps.

*Characterizing the detector dark count rate.* The measured signal ($I_M$) in a single photon detector is a sum of sample signal ($I_S$) and detector dark counts ($I_D$). To take into account the background level due to detector dark counts in the measured second-order correlation histogram, $g^{(2)}(\tau) = \langle :I_M(t+\tau)I_M(t): \rangle / \langle I_M(t) \rangle^2$, we individually measured the dark count related signal in $g^{(2)}(\tau)$. Apart from the denominator for normalization in $g^{(2)}(\tau)$, the numerator is expressed by $\langle :I_S(t+\tau)I_S(t): \rangle + 2\langle :I_M(t+\tau)I_D(t): \rangle - \langle :I_D(t+\tau) \cdot I_D(t): \rangle$. We can ignore the third term when $I_M \gg I_D$ and quantify $\langle :I_M(t+\tau)I_D(t): \rangle$ in the second term by sending the signal from the sample onto one detector while the other detector measures only the detector dark counts.

*Estimation of the Collection efficiency.* In order to estimate the collection efficiency we excited the dots with a pulsed laser and compared the photon counting rate of the dot emission at one of the detectors to the laser repetition rate of 5 MHz. Dots A and B gave a maximum single photon counting rate of 6±0.2 kcounts/s and 7.8±0.2 kcounts/s. Dividing by the system efficiency of 1.6%, which includes the transmission efficiency of optics (41%) and spectrometer (42%), the coupling efficiency to the fiber (48%), and the detector quantum efficiency (20%), we estimate the collection efficiency of 8% and 10% for dots A and B respectively.

*Theoretical model for two-photon interference measurements.* Our modified Hong-Ou-Mandel scheme measures the coincidence counts when two photons leave at the same side of a 50:50 beamsplitter, and the second order correlation functions for the two-photon interference effect between two emitters with orthogonal ($g_\perp^{(2)}(t)$) and parallel ($g_\parallel^{(2)}(t)$) polarized photons are given by[47-49]

$$g_\perp^{(2)}(t) = \frac{g_A^{(2)}(t) + g_B^{(2)}(t) + 2}{4} \quad (1)$$

$$g_\parallel^{(2)}(t) = \frac{g_A^{(2)}(t) + g_B^{(2)}(t) + 2\left(1 + V\rho_A\rho_B e^{-\frac{2|t|}{\tau_c}}\right)}{4} \quad (2)$$

where $g_{A,B}^{(2)}(t)$ and $\rho_{A,B} = 1 - \frac{B_{A,B}}{S_{A,B}}$ are the second-order auto-correlation and the ratio of the sample signal to total signal of dots A and B. $V$ and $\tau_c$ are fitted parameters denoting the wavefunction overlap of two photons and the coherence time of dots A and B. The first two terms in Eq 1,2 represent the probability of coincident counts due to multi photon events of the source itself or background emission, while the third term describes the probability of coincident counts originating from two-photon interference between two photons at opposite inputs. The sources showed a negligible detuning ($\Delta\omega$~0), and we balanced the emission intensities of two dots. We independently measured $g_{A,B}^{(2)}(t)$ and $\rho_{A,B}$=0.91 and 0.94 for dots A and B with continuous excitation and inserted those values to Eq 1,2. We convolved the equations with a Gaussian function that accounts for the limited time resolution of the detectors. With fitted parameters of $V$=0.96 and $\tau_c$=115 ps, Eq. 1,2 fit the measured data well.

AUTHOR INFORMATION

**Corresponding Author**

*E-mail: edowaks@umd.edu

ACKNOWLEDGMENT

The authors would like to acknowledge support from the Laboratory for Telecommunication Sciences, the DARPA QUINESS program (grant number W31P4Q1410003), and the Physics Frontier Center at the Joint Quantum Institute.